\documentclass[]{spie}  

\usepackage{amsmath,amsfonts,amssymb}
\usepackage{graphicx}
\usepackage{cite} 
\usepackage{epsfig}
\usepackage{amsmath}
\usepackage{nccmath}
\usepackage{amssymb}
\usepackage{mwe}
\usepackage{acro}
\usepackage{amssymb}
\usepackage{xcolor,colortbl}
\usepackage{tabularx}
\usepackage{relsize}
\usepackage{pifont}
\usepackage{booktabs} 
\usepackage{multirow}
\usepackage{multicol}
\usepackage{adjustbox}
\usepackage{float}
\usepackage{graphicx}
\usepackage{makecell}
\usepackage{tabu}
\usepackage[colorlinks=true, allcolors=blue]{hyperref}
\usepackage[capitalize]{cleveref}

\usepackage{algorithm}
\usepackage{algorithmic}

\title{Expanding Training Data for Endoscopic Phenotyping of Eosinophilic Esophagitis}

\author[a]{Juming Xiong}
\author[d]{Hou Xiong}
\author[b]{Quan Liu}
\author[b]{Ruining Deng}
\author[c]{Regina N Tyree}
\author[c]{Girish Hiremath}
\author[a,b]{Yuankai Huo}

\affil[a]{Department of Electrical and Computer Engineering, Vanderbilt University, Nashville, TN, USA}
\affil[b]{Department of Computer Science, Vanderbilt University, Nashville, TN, USA}
\affil[c]{Division of Pediatric Gastroenterology, Hepatology, and Nutrition, Vanderbilt University Medical Center, Nashville, TN, USA}
\affil[d]{Department of Physics, University of California, Santa Barbara, Santa Barbara, CA, USA}

\authorinfo{Corresponding author: Yuankai Huo: E-mail: yuankai.huo@vanderbilt.edu}

\begin{document} 
\maketitle

\begin{abstract}
Eosinophilic esophagitis (EoE) is a chronic esophageal disorder marked by eosinophil-dominated inflammation. Diagnosing EoE usually involves endoscopic inspection of the esophageal mucosa and obtaining esophageal biopsies for histologic confirmation. Recent advances have seen AI-assisted endoscopic imaging, guided by the EREFS system, emerge as a potential alternative to reduce reliance on invasive histological assessments. Despite these advancements, significant challenges persist due to the limited availability of data for training AI models $-$ a common issue even in the development of AI for more prevalent diseases. This study seeks to improve the performance of deep learning-based EoE phenotype classification by augmenting our training data with a diverse set of images from online platforms, public datasets, and electronic textbooks increasing our dataset from 435 to 7050 images. We utilized the Data-efficient Image Transformer for image classification and incorporated attention map visualizations to boost interpretability. The findings show that our expanded dataset and model enhancements markedly improve diagnostic accuracy, robustness, and comprehensive analysis, enhancing patient outcomes.

\end{abstract}


\section{INTRODUCTION}
\label{sec:intro}  
EoE has been increasing in recent decades, with approximately five to ten new cases per 100,000 inhabitants annually for both children and adults. This rise has contributed to the growing prevalence, which was 15.4 per 100,000 inhabitants before 2007 and has reached 63.2 per 100,000 inhabitants since 2017\cite{arias2020epidemiology}. Clinically, it manifests as symptoms associated with esophageal dysfunction and is characterized by an eosinophil-predominant inflammation defined as $\geq$ 15 eosinophils (Eos) (peak count) per high power field (HPF)~\cite{LIACOURAS20113}. Deep learning methods to detect and enumerate Eos has achieved very good results~\cite{xiong2024deep, liu2024eosinophils,xiong2024circlerepresentationmedicalinstance}. However, this requires biopsy and tens of thousands of annotations. Recently, endoscopic images based on the Eosinophilic Esophagitis Endoscopic Reference Score (EREFS) system have been used in conjunction with histologic findings to measure outcomes in pediatric patients and to evaluate treatment approaches for children with EoE~\cite{Wechsler2017EosinophilicER}. 

\begin{figure*}[t]
\begin{center}
\includegraphics[width=1\linewidth]{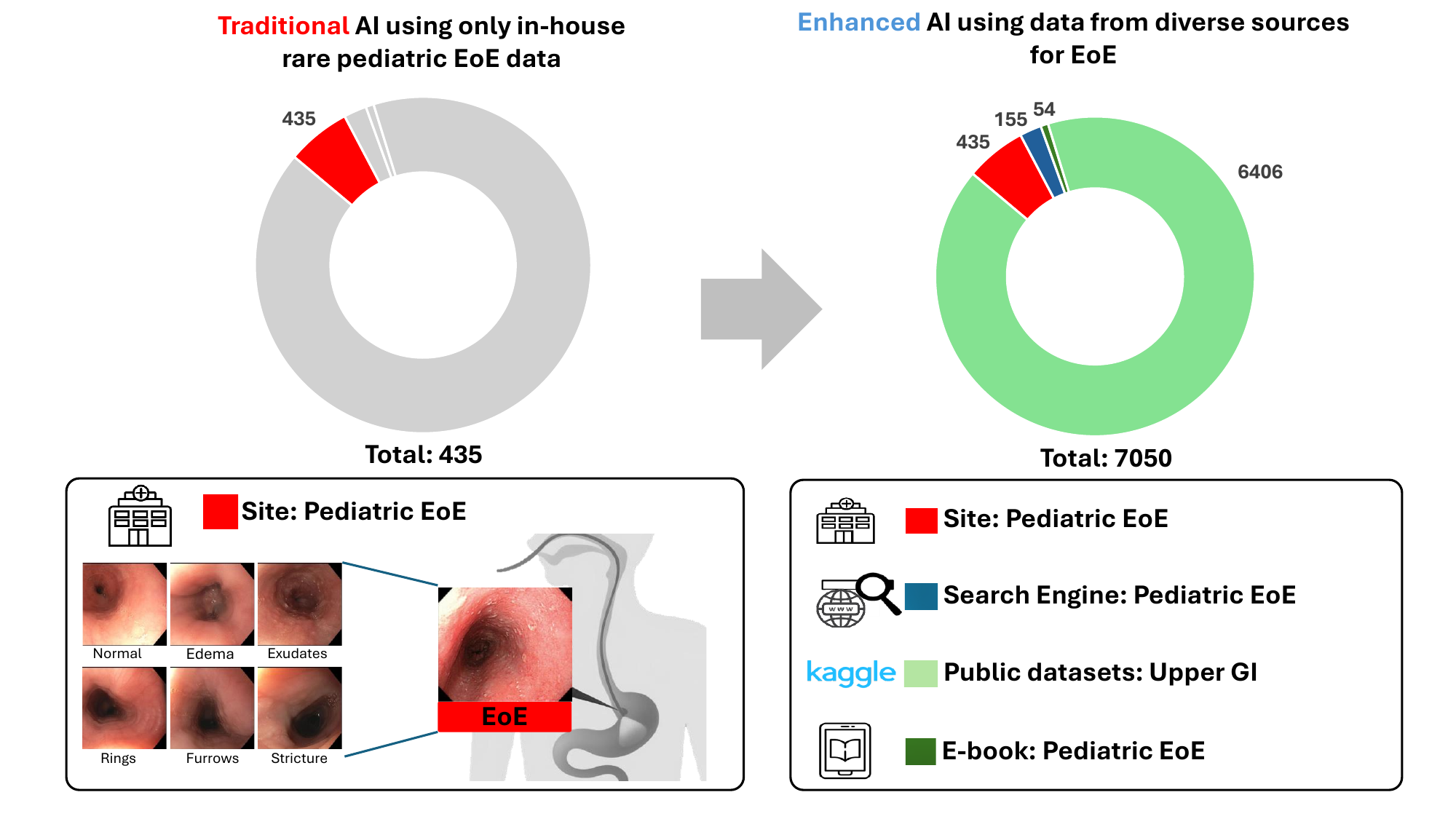}
\end{center}
\caption{This figure shows that we expanded our training set from 435 to 7050 images for training EoE deep learning models. The left panel shows that the traditional AI model only uses rare pediatric EoE data. The right panel shows that we employ a wider range of data from different sources.}
\label{fig:overview}
\end{figure*}

However, patient privacy issues contribute to the scarcity of publicly available data, thereby constraining the dataset available for training AI models in pediatric EoE. In this paper, we mined images from different sources to expand our dataset and enhance our deep-learning image classification model as shown in Figure~\ref{fig:overview}.

\section{Method}
\subsection{EREFS System and Related Esophageal Conditions}

The EREFS system provides a standardized approach for documenting endoscopic findings in patients diagnosed with EoE~\cite{article1}. It categorizes the endoscopic appearance into five key features: Edema, which involves the swelling of the esophageal mucosa leading to a loss of the vascular pattern; Rings, characterized by visible concentric circles or 'trachealization' within the esophagus; Exudates, which are white plaques or specks indicative of eosinophilic abscesses; Furrows, presenting as vertical lines within the esophageal mucosa; and Strictures, which are areas where the esophageal lumen narrows, potentially in a focal or diffuse manner.

In addition to EoE, several related esophageal conditions are noteworthy~\cite{article2}. Esophagitis refers to inflammation of the esophagus, which can arise from various factors including acid reflux (GERD), infections, and allergies like EoE, often causing symptoms like heartburn, chest pain, and difficulty swallowing. The Z-Line marks the gastroesophageal junction where squamous epithelium transitions to columnar epithelium. Barrett's Esophagus is a serious condition where this normal squamous lining is replaced with columnar epithelium due to prolonged acid exposure from GERD, increasing the risk of developing esophageal adenocarcinoma. The pylorus controls the flow of stomach contents into the duodenum. Additionally, the retroflex stomach maneuver allows endoscopic examination of the gastric cardia, fundus, and lower esophagus by bending the scope backward within the stomach.













\begin{figure*}[t]
\begin{center}
\includegraphics[width=1\linewidth]{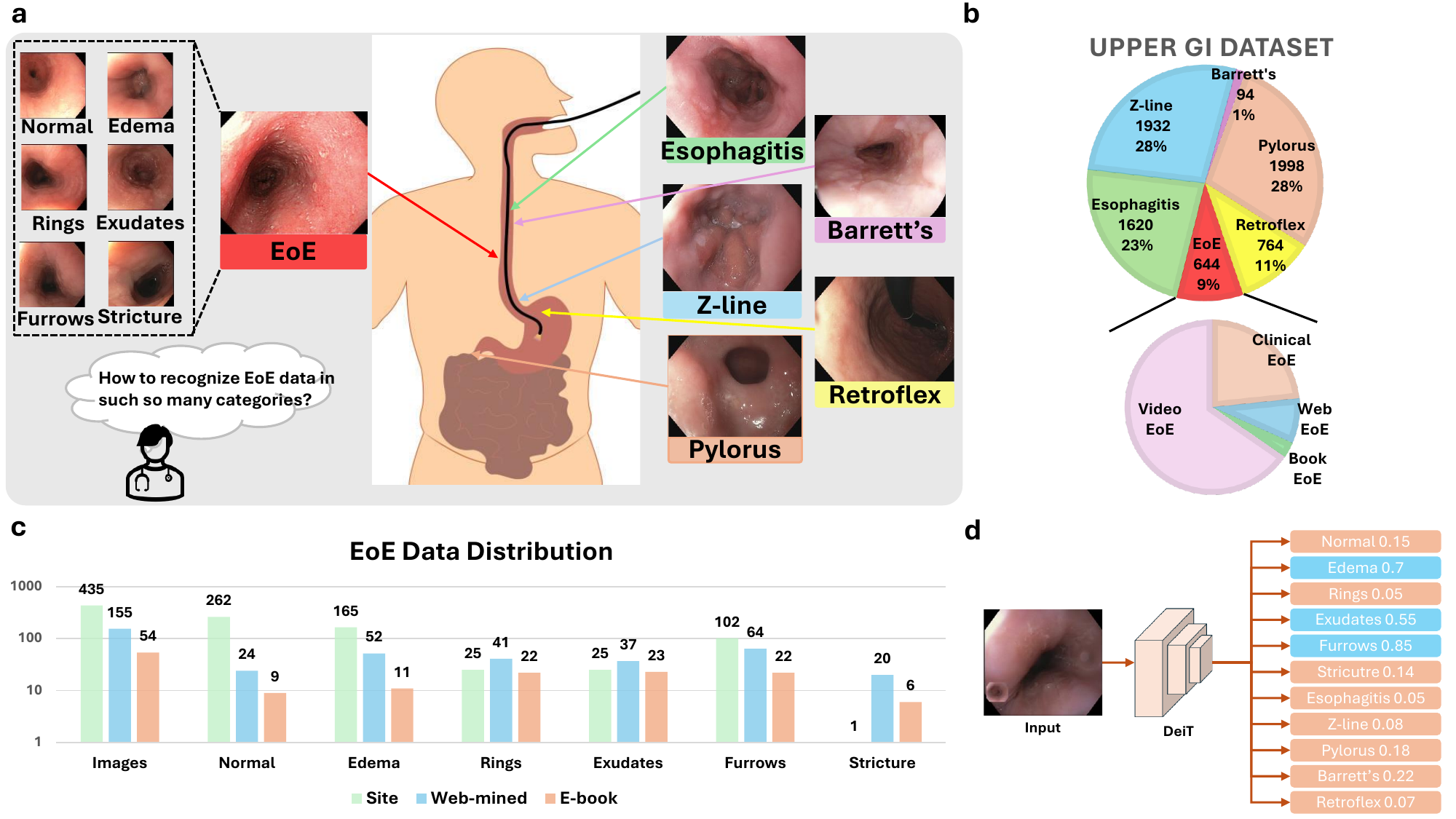}
\end{center}
\caption{This figure provides an overview of the learning framework. (a) illustrates the image classes, which include six EoE classes and five non-EoE classes, along with their approximate locations in the upper GI tract. (b) depicts the proportion of each class within the entire upper GI dataset. (c) displays the distribution of the EoE dataset after mining or collection. Finally, (d) shows how the DeiT AI model classifies these images.}
\label{fig:Method}
\end{figure*}

\subsection{Image Mining from Search Engine and E-books}
Inspired by prior studies\cite{yao2022compound,Huang2023AVF}, we mined data from various online sources, including Google, the NIH open-I public dataset, social media platforms like Twitter, Facebook, LinkedIn, and other sources to determine if website data could help the model learn diverse features and improve performance. Given the numerous phenotypes in the upper GI tract, we expanded our dataset by incorporating the Kavsir dataset~\cite{Pogorelov:2017:KMI:3083187.3083212}, which includes a range of conditions such as GERD, Barrett's esophagus, and infectious esophagitis. Specifically, we used the Icrawler open-source data mining toolbox to crawl images from Google. Meanwhile, we utilized the NIH open-I API to search for and download images. For E-book EoE data, we searched for related books using Google Books and Vanderbilt library resources and downloaded the images, which have corresponding captions indicating their class. For social media images, such as those on Twitter, we searched for keywords to find relevant images. We initially screened these images using a basic AI model to remove irrelevant ones, followed by further screening and verification by doctors except E-book data because they already have high-quality labels.

This expansion provides several potential advantages. Firstly, it might allow our AI model to learn and distinguish the varied presentations of different upper GI diseases, improving diagnostic accuracy for EoE and reducing misclassification. Secondly, it would enhance the model's ability to perform comprehensive analyses, identifying potential comorbidities and providing a more holistic view of the patient's health. Lastly, including additional phenotypes increases our model's robustness and generalizability, ensuring it performs well across different clinical settings and patient populations as shown in Figure~\ref{fig:Method}. 


\subsection{Image Classification}
Since the Vision Transformer (ViT) introduction by Dosovitskiy \textit{et al}~\cite{DBLP:journals/corr/abs-2010-11929}., ViT has been widely adopted for image classification, leveraging the power of self-attention mechanisms originally developed for natural language processing. This model's ability to handle image data by treating image patches as tokens has opened new avenues for applying transformer-based architectures to visual tasks. The Data-efficient Image Transformer (DeiT) represents a notable advancement in computer vision, building on the strengths of the ViT while addressing its limitations in data efficiency and training requirements. DeiT retains the ability to model long-range dependencies within images through self-attention mechanisms, capturing complex patterns often missed by conventional convolutional neural networks (CNNs). Its distillation approach, involving a CNN-based teacher model and a learnable distillation token, allows effective training with smaller datasets.

\subsection{Aattention-based Visualization}
We utilized Gradient Attention Rollout to visualize the attention map. This method is based on aggregating attention weights across multiple layers in a neural network to identify the most influential parts of the input data. Specifically, the attention map A is computed as follows:
\begin{equation}
A = \sum_{l=1}^{L} \alpha_l \cdot A_l
\end{equation}

\noindent where $\alpha_l$ represents the attention weights from the l-th layer, and $A_l$ is the attention map at that layer. The benefit of this method lies in its ability to provide insights into the model's decision-making process by highlighting which parts of the input are most critical for the output. This visualization technique enhances interpretability and allows researchers to understand and trust the model's predictions better, thereby facilitating the identification of potential biases or areas for improvement.

\section{Data and Experiments}

\subsection{Data}

435 endoscopy pediatric EoE images were collected from Vanderbilt Univesity Medical Center (Site), which include 262 normal labels, 165 edema labels, 25 rings labels, 25 exudates labels, 102 furrows labels and 1 stricture label. Then, we mined 155 images from online sources which included 24 normal labels 52 edema labels, 41 rings labels, 37 exudates labels, 64 furrows labels, and 20 stricture labels. Next, we also searched electrical pediatric EoE text books for high-quality testing data. It had 54 images, including 9 normal labels, 11 edema labels, 22 rings labels, 23 exudates labels, 22 furrows labels and 5 stricture labels. Since images in the electrical textbook have high-quality annotation, all images with annotation are only used in validation and test datasets. Specifically, it included 579 normal labels, 610 edema labels, 290 rings labels, 140 exudates labels, 241 furrows labels, and 19 stricture labels. Additionally, to simulate the common condition in the upper gastrointestinal (GI), we also used other conditions in upper GI as the negative control, which totally includes 6406 images, 1620 esophagitis images, 1020 z-line images, 94 barretts images, 1998 pylorus images and 764 retroflex stomach images~\cite{Pogorelov:2017:KMI:3083187.3083212}. The training dataset, validation dataset and testing dataset are approximately as the ratio of 7:1:2 respectfully. The detailed image and level are distributed as the following Table~\ref{tab:data_summary} and Table~\ref{tab:upper_gi_summary}.

\begin{table}[ht]
\centering
\begin{tabular}{lrrrrrrr}
\toprule
\textbf{Source} & \textbf{Images} & \textbf{Normal} & \textbf{Edema} & \textbf{Rings} & \textbf{Exudates} & \textbf{Furrows} & \textbf{Stricture} \\ 
\midrule
\multicolumn{8}{c}{\textbf{Training Data}} \\ 
\midrule
Site      & 304             & 183             & 113            & 17             & 16                & 68               & 0                 \\ 
Web-mined       & 108             & 19              & 38             & 27             & 25                & 42               & 13                \\ 
\textbf{Total}  & \textbf{412}   & \textbf{202}    & \textbf{151}   & \textbf{44}   & \textbf{41}      & \textbf{110}     & \textbf{13}       \\ 
\midrule
\multicolumn{8}{c}{\textbf{Validation Data}} \\ 
\midrule
Site      & 44              & 31              & 13             & 4              & 1                 & 10               & 1                 \\ 
Web-mined       & 16              & 3               & 3              & 5              & 5                 & 7                & 2                 \\ 
E-book          & 20              & 3               & 6              & 8              & 8                 & 9                & 3                 \\ 
\textbf{Total}  & \textbf{80}     & \textbf{37}     & \textbf{22}    & \textbf{17}    & \textbf{14}       & \textbf{26}      & \textbf{6}        \\ 
\midrule
\multicolumn{8}{c}{\textbf{Testing Data}} \\ 
\midrule
Site      & 87              & 48              & 39             & 4              & 8                 & 24               & 0                 \\ 
Web-mined       & 31              & 2               & 11             & 9              & 7                 & 15               & 5                 \\ 
E-book          & 34              & 6               & 5              & 14             & 15                & 13               & 3                 \\ 
\textbf{Total}  & \textbf{152}    & \textbf{56}     & \textbf{55}    & \textbf{27}    & \textbf{30}       & \textbf{52}      & \textbf{8}        \\ 
\midrule
\multicolumn{8}{c}{\textbf{Combined Total}} \\ 
\midrule
Site      & 435             & 262             & 165            & 25             & 25                & 102              & 1                 \\ 
Web-mined       & 155             & 24              & 52             & 41             & 37                & 64               & 20                \\ 
E-book          & 54              & 9               & 11             & 22             & 23                & 22               & 6                 \\ 
\textbf{Total}  & \textbf{644}    & \textbf{295}    & \textbf{228}   & \textbf{88}    & \textbf{85}       & \textbf{188}     & \textbf{27}       \\ 
\bottomrule
\end{tabular}
\caption{Data Summary for Training, Validation, and Testing for EoE}
\label{tab:data_summary}
\end{table}

\begin{table}[ht]
\centering
\begin{tabular}{lrrrrrr}
\toprule
\textbf{Dataset} & \textbf{Images} & \textbf{Esophagitis} & \textbf{Z-line} & \textbf{Barrett's} & \textbf{Pylorus} & \textbf{Retroflex Stomach} \\ 
\midrule
\textbf{Train}  & 4481            & 1133                 & 1351            & 65                 & 1398             & 534                     \\ 
\textbf{Val}    & 644             & 163                  & 194             & 10                 & 200              & 77                      \\ 
\textbf{Test}   & 1281            & 324                  & 387             & 19                 & 400              & 153                     \\ 
\midrule
\textbf{Total}  & \textbf{6406}   & \textbf{1620}        & \textbf{1932}   & \textbf{94}        & \textbf{1998}    & \textbf{764}            \\ 
\bottomrule
\end{tabular}
\caption{Data Summary for Upper GI Dataset}
\label{tab:upper_gi_summary}
\end{table}

\subsection{Experiment}

\subsubsection{Training Detail}
The network underwent training with fine-tuned parameters derived from the validation set, encompassing a learning rate of 0.001. The input image resolution was configured as 224$\times$224$\times$3. For dataset augmentation, we utilized randomly flips the images horizontally and applies random rotations up to 90 degrees. Gaussian blur is applied with a probability of 100$\%$, followed by color jittering to adjust brightness, contrast, saturation, and hue slightly. 

The implementation of the network was carried out using Python version 3.8 and PyTorch version 2.0.1, utilizing CUDA version 11.7 for GPU acceleration. The experiments were performed on an NVIDIA RTX A6000 GPU with 48GB memory, enabling efficient processing and training of the model.

\subsubsection{Evaluation metric}
Since our task involves multi-label image classification, we adopted a one-hot encoding scheme for labeling the data. This approach allows us to represent each image with a binary vector, where each element indicates the presence or absence of a specific label. By doing so, we can effectively manage the complexity of assigning multiple labels to a single image. To evaluate the performance of our model, we utilized the F1 score, a metric that balances precision and recall. The F1 score is particularly suitable for multi-label tasks as it provides a comprehensive measure of the model's ability to correctly identify each label while minimizing both false positives and false negatives.

\section{Results}
For the EoE-related categories, the performance metrics across different datasets reveal that the model using combination of Site and Web-mined, generally achieves the highest scores. Specifically, this dataset exhibits the highest values for EoE-related metrics such as EoE (60.00\%), Exudates (43.90\%), Furrows (58.95\%), and Stricture (20.00\%). In contrast, the model only using the Site dataset shows relatively lower performance across most EoE-related metrics.

In the Non-EoE categories, the performance metrics do not show obvious differences across datasets. The model only using the Site dataset demonstrates competitive performance in Non-EoE, Esophagitis, Z-line, Barretts, Pylorus, and Retroflex Stomach, with a slight improvement observed when combining the Site and Web-mined datasets as shown in Table~\ref{tab:result}. 

\begin{figure*}[t]
\begin{center}
\includegraphics[width=0.8\linewidth]{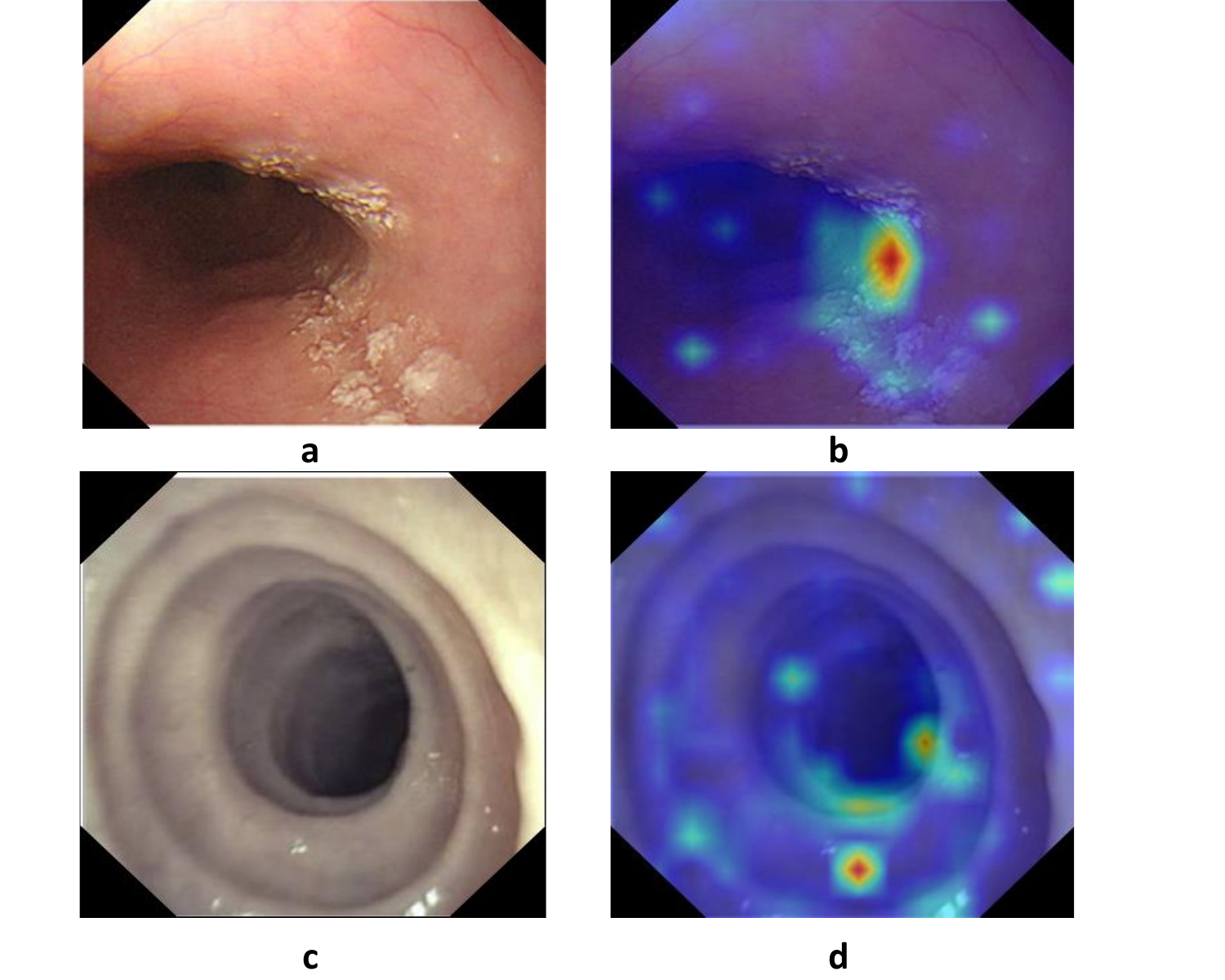}
\end{center}
\caption{This figure shows the endoscopy EoE image and corresponding attention map. Image a shows an endoscopic image with label ``exudate", whose attention map is presented in image b. Image c presents an endoscopic image with ``rings", whose attention map visualization}
\label{fig:visualize}
\end{figure*}

The attention map highlight specific features the model attends to when diagnosing different conditions. For instance, part (a) shows an endoscopic image with exudates, while part (b) reveals the corresponding attention map where the model identifies and highlights the white exudates effectively. Similarly, part (c) and part (d) illustrate an image with rings and the corresponding attention map, respectively. These visualizations validate that the model accurately focuses on clinically relevant features, enhancing interpretability and trust in the model's diagnostic capabilities as shown in Figure~\ref{fig:visualize}.

\begin{table}[ht]
\centering
\begin{tabular}{cc|rrrrrrr}
\toprule
\multicolumn{2}{c|}{\textbf{Data Sources}} & \multicolumn{7}{c}{\textbf{Classes}} \\ 
\cmidrule(r){1-2} \cmidrule(r){3-9}
\textbf{Site} & \textbf{Web-mined} & \textbf{EoE} & \textbf{Normal} & \textbf{Edema} & \textbf{Rings} & \textbf{Exudates} & \textbf{Furrows} & \textbf{Stricture} \\ 
\midrule
\checkmark &  & 57.79 & \textbf{75.44} & \textbf{57.63} & 55.81 & 41.03 & 47.37 & 00.00 \\ 
\checkmark & \checkmark &  \textbf{60.00} & 71.70 & 55.05 & \textbf{67.80} & \textbf{43.90} & \textbf{58.95} & \textbf{20.00} \\ 
\bottomrule
\end{tabular}

\begin{tabular}{cc|rrrrrr}
\toprule
\multicolumn{2}{c|}{\textbf{Data Sources}} & \multicolumn{6}{c}{\textbf{Classes}} \\ 
\cmidrule(r){1-2} \cmidrule(r){3-8}
\textbf{Site} & \textbf{Web-mined} & \textbf{Non EoE} & \textbf{Esophagitis} & \textbf{Z-line} & \textbf{Barretts} & \textbf{Pylorus} & \textbf{Retroflex Stomach} \\ 
\midrule
\checkmark &  &  92.37 & 84.75 & 91.95 & 22.22 & 99.13 & 100.00 \\ 
\checkmark & \checkmark &  \textbf{93.77} & \textbf{88.28} & \textbf{92.49} & \textbf{25.00} & \textbf{99.88} & 100.00 \\ 
\bottomrule
\end{tabular}
\caption{Test performance metrics for the models using different data sources}
\label{tab:result}
\end{table}

\section{Conclusion}

In this paper, we have introduced an advanced endoscopy image classification model specifically designed for the accurate detection and analysis of EoE. Our proposed model enhances the precision of EoE identification and supports a comprehensive evaluation of associated conditions, ultimately contributing to more informed clinical decision-making and improved patient outcomes. A key aspect of our approach involves the integration of web-mined EoE images, which provides a diverse and extensive dataset. This rich variety of images facilitates the model's ability to recognize a wide range of disease presentations, thereby greatly improving diagnostic accuracy and model robustness.

The amalgamation of these distinct datasets enhances the model's performance by enabling it to generalize more effectively across different patient demographics. This dual-dataset approach underscores the model's capability to deliver reliable and precise classifications, thereby advancing the overall efficacy of EoE diagnosis and management.

\section{ACKNOWLEDGMENTS}       
This research was supported by NIH R01DK135597(Huo), DoD HT9425-23-1-0003(HCY), NIH NIDDK DK56942(ABF). This work was also supported by Vanderbilt Seed Success Grant, Vanderbilt Discovery Grant, and VISE Seed Grant. This project was supported by The Leona M. and Harry B. Helmsley Charitable Trust grant G-1903-03793 and G-2103-05128. This research was also supported by NIH grants R01EB033385, R01DK132338, REB017230, R01MH125931, and NSF 2040462. We extend gratitude to NVIDIA for their support by means of the NVIDIA hardware grant.

\bibliography{main} 

\begin{thebibliography}{10}

\bibitem{arias2020epidemiology}
Arias, {\'A}. and Lucendo, A.~J., ``Epidemiology and risk factors for eosinophilic esophagitis: lessons for clinicians,'' {\em Expert review of gastroenterology \& hepatology}~{\bf 14}(11),  1069--1082 (2020).

\bibitem{LIACOURAS20113}
Liacouras, C.~A., Furuta, G.~T., Hirano, I., Atkins, D., Attwood, S.~E., Bonis, P.~A., Burks, A.~W., Chehade, M., Collins, M.~H., Dellon, E.~S., Dohil, R., Falk, G.~W., Gonsalves, N., Gupta, S.~K., Katzka, D.~A., Lucendo, A.~J., Markowitz, J.~E., Noel, R.~J., Odze, R.~D., Putnam, P.~E., Richter, J.~E., Romero, Y., Ruchelli, E., Sampson, H.~A., Schoepfer, A., Shaheen, N.~J., Sicherer, S.~H., Spechler, S., Spergel, J.~M., Straumann, A., Wershil, B.~K., Rothenberg, M.~E., and Aceves, S.~S., ``Eosinophilic esophagitis: Updated consensus recommendations for children and adults,'' {\em Journal of Allergy and Clinical Immunology}  (2011).

\bibitem{xiong2024deep}
Xiong, J., Liu, Y., Deng, R., Tyree, R.~N., Correa, H., Hiremath, G., Wang, Y., and Huo, Y., ``Deep learning-based open source toolkit for eosinophil detection in pediatric eosinophilic esophagitis,'' in [{\em Medical Imaging 2024: Digital and Computational Pathology}{\nolinebreak\hspace{0.1em}]},   {\bf 12933},  231--237, SPIE (2024).

\bibitem{liu2024eosinophils}
Liu, Y., Deng, R., Xiong, J., Tyree, R.~N., Correa, H., Hiremath, G., Wang, Y., and Huo, Y., ``Eosinophils instance object segmentation on whole slide imaging using multi-label circle representation,'' in [{\em Medical Imaging 2024: Digital and Computational Pathology}{\nolinebreak\hspace{0.1em}]},   {\bf 12933},  114--132, SPIE (2024).

\bibitem{xiong2024circlerepresentationmedicalinstance}
Xiong, J., Nguyen, E.~H., Liu, Y., Deng, R., Tyree, R.~N., Correa, H., Hiremath, G., Wang, Y., Yang, H., Fogo, A.~B., and Huo, Y., ``Circle representation for medical instance object segmentation,'' (2024).

\bibitem{Wechsler2017EosinophilicER}
Wechsler, J.~B., Bolton, S.~M., Amsden, K., Wershil, B.~K., Hirano, I., and Kagalwalla, A.~F., ``Eosinophilic esophagitis reference score accurately identifies disease activity and treatment effects in children,'' {\em Clinical Gastroenterology \& Hepatology}~{\bf 16},  1056–1063 (2017).

\bibitem{article1}
Hirano, I., Moy, N., Heckman, M., Thomas, C., Gonsalves, N., and Achem, S., ``Endoscopic assessment of the oesophageal features of eosinophilic oesophagitis: Validation of a novel classification and grading system,'' {\em Gut}~{\bf 62} (05 2012).

\bibitem{article2}
Carlson, D., Hinchcliff, M., and Pandolfino, J., ``Advances in the evaluation and management of esophageal disease of systemic sclerosis,'' {\em Current rheumatology reports}~{\bf 17},  475 (01 2015).

\bibitem{yao2022compound}
Yao, T., Qu, C., Long, J., Liu, Q., Deng, R., Tian, Y., Xu, J., Jha, A., Asad, Z., Bao, S., et~al., ``Compound figure separation of biomedical images: Mining large datasets for self-supervised learning,'' {\em The journal of machine learning for biomedical imaging}~{\bf 1} (2022).

\bibitem{Huang2023AVF}
Huang, Z., Bianchi, F., Yuksekgonul, M., Montine, T.~J., and Zou, J.~Y., ``A visual–language foundation model for pathology image analysis using medical twitter,'' {\em Nature Medicine}~{\bf 29},  2307--2316 (2023).

\bibitem{Pogorelov:2017:KMI:3083187.3083212}
Pogorelov, K., Randel, K.~R., Griwodz, C., Eskeland, S.~L., de~Lange, T., Johansen, D., Spampinato, C., Dang-Nguyen, D.-T., Lux, M., Schmidt, P.~T., Riegler, M., and Halvorsen, P., ``Kvasir: A multi-class image dataset for computer aided gastrointestinal disease detection,'' in [{\em Proceedings of the 8th ACM on Multimedia Systems Conference}{\nolinebreak\hspace{0.1em}]},  {\em MMSys'17},  164--169, ACM, New York, NY, USA (2017).

\bibitem{DBLP:journals/corr/abs-2010-11929}
Dosovitskiy, A., Beyer, L., Kolesnikov, A., Weissenborn, D., Zhai, X., Unterthiner, T., Dehghani, M., Minderer, M., Heigold, G., Gelly, S., Uszkoreit, J., and Houlsby, N., ``An image is worth 16x16 words: Transformers for image recognition at scale,'' {\em CoRR}~{\bf abs/2010.11929} (2020).

\end{thebibliography}
\bibliographystyle{spiebib} 

\end{document}